\documentclass[11pts,fleqn]{article}
\pdfoutput=1
\usepackage{jcappub}
\usepackage[T1]{fontenc} 
\usepackage[german,british]{babel}
\usepackage{graphicx}
\usepackage{amssymb}
\usepackage{latexsym}

\newcommand{\be}{\begin{equation}}
\newcommand{\ee}{\end{equation}}
\newcommand{\dd}{{\rm d}}
\newcommand{\lp}{\left(}
\newcommand{\rp}{\right)}

\newcommand{\mP}{\mathcal{P}}
\newcommand{\rP}{{\rm P}}
\newcommand{\mJ}{\mathcal{J}}

\title{\textbf{Counts of galaxy clusters as cosmological probes: the impact of baryonic physics}}
\author{Andr\'es Balaguera-Antol\'{\i}nez,}
\author{Cristiano Porciani}
\affiliation{Argelander Institute f\"ur Astronomie, Auf dem H\"ugel 71, D-53121 Bonn, Germany}
\emailAdd{abalan@astro.uni-bonn.de}
\emailAdd{porciani@astro.uni-bonn.de}

\abstract{
  The halo mass function from N-body simulations of collisionless matter
  is generally used to retrieve cosmological parameters from observed counts of galaxy clusters.
  This neglects the observational fact that the baryonic mass fraction in clusters
  is a random variable that, on average, increases with the total mass (within an overdensity of $500$).
  Considering a mock catalog that includes tens of thousands of galaxy clusters,
  as expected from the forthcoming generation of surveys, we show that the effect of a 
  varying baryonic mass fraction will be observable with high statistical significance. 
  The net effect is a change in the overall normalization of the cluster mass
  function and a milder modification of its shape. Our results indicate the necessity 
  of taking into account baryonic corrections to the mass function if one wants
  to obtain unbiased estimates of the cosmological parameters from data
  of this quality. We introduce the formalism necessary to accomplish this goal. 
  Our discussion is based on the conditional probability of finding a given value of 
  the baryonic mass fraction for clusters of fixed total mass. 
  Finally, we show  that combining information from the cluster counts with measurements
  of the baryonic mass fraction in a small subsample of clusters (including only a few tens 
  of objects) will nearly optimally constrain the cosmological parameters.
}


\notoc
\begin{document}
\maketitle
\flushbottom

\section{Introduction}
Galaxy clusters are powerful cosmological probes \cite{vik,cun1,mantzI,rapetti}
which are expected to be associated with massive haloes of dark matter. 
To date, 
two different procedures have been employed to extract cosmological
information from large samples of galaxy clusters. 
The simplest one is to measure their observed
number density as a function of mass and/or redshift and compare it with
theoretical models for the halo abundance which are calibrated against
N-body simulations of collisionless matter \cite{vik}. The second
option is to use their baryonic content within the virial radius as a sort of 
standard ruler \cite{2003A&A...398..879E,2008MNRAS.383..879A}. 
This assumes that the baryon fraction in clusters
coincides with the universal value \cite{1993Natur.366..429W} and does not scatter from object to object.
The recent analysis by \cite{2013arXiv1301.0624E} has shown that this is a good assumption for
relaxed, cool-core clusters.
In this paper, we revisit the first approach and show that some modifications
to the standard procedure will be needed when the forthcoming generation
of cluster surveys will become available. Therefore, the second method will
not be discussed any further. 
  
The key theoretical quantity to analyze 
cluster-count experiments 
is the halo mass function $n_{\rm h}(M,z)$ which is defined so that $n_{\rm h}(M,z)\,\dd M$, 
gives the number of dark-matter haloes (at redshift $z$) with mass between 
$M$ and $M+\dd M$ per unit comoving volume. 
The calibration of the halo mass function using N-body simulations
has been pushed to a 
precision of $\sim 5\%$ \cite{jenkins,warren,tinker,pillepich_nm} and can be even further improved \cite{reed}. However, 
this precision is somewhat illusory for direct applications to galaxy clusters. In fact, numerical 
simulations that, more realistically, also consider a baryonic component on top of the dark matter 
\cite{stanek_bar,McCarthy,2012MNRAS.423.2279C,daalen,rasia} have shown that gas physics can alter cluster masses in a systematic way (i.e. in one specific
direction and not randomly) by 
up to $\sim 10\%$ \cite{2012MNRAS.423.2279C} with respect to pure N-body simulations. Consequently, given the steep slope of the halo mass
function at cluster scales, the cluster 
abundance at fixed mass changes by $\sim 10-20\%$ \cite{stanek_bar}. This is larger than the forecasted statistical uncertainties ($\sim 1-10\%$) for the forthcoming generation of cluster surveys.

Regrettably, due to the large uncertainties in 
modeling baryon physics (i.e. radiative cooling, non-thermal pressure 
support,
star formation, and different feedback mechanisms in the 
presence of stars and active galactic nuclei), current gas simulations can only provide order-of-magnitude 
approximations of these effects 
(see Section 2 for an extended discussion) and cannot provide robust
estimates for the cluster mass function.

State-of-the-art techniques to constrain cosmological parameters from cluster
counts take into consideration the effects of 
baryons by artificially enlarging the uncertainties in the parameters of the halo mass function
extracted from N-body simulations and marginalizing over them \cite{mantzI}. This is a statistical
trick to account for possible differences between the shapes of the halo and cluster mass functions. 
Such an approach is adequate for the current samples of clusters which probe 
relatively small volumes and have large error bars \cite{2012MNRAS.425.2244B}. 
Forthcoming cluster surveys will probe much larger volumes with higher 
sensitivities.
In this paper, we show that, in order to use the new data to derive
cosmological constraints that are not only
precise (i.e. small statistical errors) but also accurate 
(i.e. unbiased estimates), it will be imperative to model the baryonic effects on the cluster mass 
function.

As routinely done in X-ray and Sunyaev-Zel'dovich studies,
we define cluster and halo masses as spherical overdensities bounded by the
radius $R_{500}$ within which the mean density is  
$\Delta=500$ times the critical density of the Universe. Note the $R_{500}$
is typically 1.5-2 times smaller than the virial radius (defined as in \cite{1996MNRAS.282..263E}) and the baryon fraction
within this smaller scale, $f_{\rm b}$, is observed to be below the
cosmic value (e.g. \cite{andreon, 2013arXiv1301.0624E})
and vary systematically with the cluster mass
\cite{giodini,lagana,lin,sun}.
Following a statistical approach, we describe the effects of gas physics in
two steps. We first use the outcome of recent numerical studies, to relate
the dark-matter content of galaxy clusters with halo masses in  
N-body simulations. Then we reason in terms of the mean baryonic mass fraction 
as a function of cluster mass, $f_{\rm b}(M)$, and of the corresponding dispersion around the mean $\sigma_{\rm b}$.
We focus our attention on two cosmological parameters: 
the matter density parameter, $\Omega_{\rm m}$, and  the linear rms matter fluctuation within
a spherical top-hat window of radius $8\,h^{-1}$ Mpc, $\sigma_8$.
Considering observationally motivated guesses for the shape and amplitude of the function $f_{\rm b}(M)$ and
of the scatter $\sigma_{\rm b}$, we first quantify the bias in the estimates for $\Omega_{\rm m}$ and $\sigma_8$ caused 
by using the halo mass function as a proxy for the cluster mass function. Subsequently, we explore the option of 
using extra free parameters in the models to simultaneously determine from the
cluster counts both the cosmological parameters and the relation $f_{\rm b}(M)$.
We find that this procedure increases the uncertainties on $\Omega_{\rm m}$ and $\sigma_8$.  
Finally, we show that considering additional information on the baryonic mass fraction
from follow-up studies of a small subsample of clusters is pivotal to recover unbiased estimates for 
$\Omega_{\rm m}$ and $\sigma_8$ with small uncertainties.

We conclude that taking into account the effect of baryons on the cluster mass 
function is key to fully exploit the potential of forthcoming large-volume 
cluster surveys as cosmological probes.

We adopt a fiducial cosmological model based on a flat $\Lambda$CDM Universe with a matter density parameter of 
$\Omega_{\rm m}=0.258$, a baryon density parameter of $\Omega_{\rm b}=0.044$, a dimensionless Hubble parameter $h=0.735$ 
(in units of $100 \,{\rm km}\,{\rm s}^{-1} {\rm Mpc}^{-1}$), a linear rms mass fluctuation within $8\,h^{-1}$ Mpc 
of $\sigma_{8}=0.773$ and a scalar spectral index $n_{s}=0.954$. We use the transfer function of \cite{ehu} to compute 
the linear matter power spectrum.  
The halo mass function is calculated using the fitting formulae of \cite{tinker}. 
Logarithms are always taken after measuring the mass in units of $h^{-1} M_\odot$.

\section{Matching cluster masses to N-body simulations }

Many authors have compared the mass density profiles of clusters generated
from hydrodynamic and collisionless simulations with
identical initial conditions.
The presence of the baryons generates 
not only a change in the total mass (defined within a fixed overdensity) of a cluster 
compared to the corresponding halo in a N-body simulation, but
also a re-distribution of the dark-matter component 
\cite{2004ApJ...616...16G,2008ApJ...672...19R,2010MNRAS.405.2161D}. 
When the gas is not allowed to radiate, infalling baryons are heated up 
in shocks when they cross the virial radius and subsequently exchange energy 
with the dark matter while the cluster undergoes dynamical relaxation. 
They end up having a more extended distribution than the dark matter \cite{1998ApJ...503..569E,
1999ApJ...525..554F,2006MNRAS.365.1021E,2006AIPC..878....3G,2007MNRAS.377...41C,
2008ApJ...672...19R}. As a matter of fact, the baryon fraction within the virial radius
decreases with the cluster mass and lies below the cosmic value, $f_{\rm c}\equiv \Omega_{\rm b}/\Omega_{\rm m}$, even for the
most massive clusters (see Fig. 5 in \cite{2008ApJ...672...19R}).
The dark-matter profile in spherical shells becomes slightly more concentrated
than in N-body simulations
(see Fig. 6 in \cite{2008ApJ...672...19R}) and the total dark-matter mass within
$\Delta=500$ increases by $\sim 1 \%$ for an object of $10^{14} h^{-1} M_\odot$.
%
When, instead, gas cooling and star formation are included in simulations, 
baryons
dissipate energy radiatively and condense in the central regions pulling the
dark matter along. Orbits of substructures are also altered \cite{2012MNRAS.423.2279C}.
If the energy feedback due to star formation is weak, 
the net effect is that the concentration of the
dark matter increases \cite{2008ApJ...672...19R,2012MNRAS.423.2279C}. In this case, clusters exhibit 
dark-matter (and total) masses within $R_{500}$
which are a few per cent higher than in the 
corresponding N-body simulations. Note, however, that the baryon fractions of the simulated clusters 
within the virial radius systematically lie 1-5\% above the universal value. Stellar profiles also deviate substantially from
the observed ones \cite{2010MNRAS.405.2161D,2012MNRAS.423.2279C}. Observations \cite{giodini,lagana,lin,sun} have shown that the baryon fraction within $R_{500}$ lies 
below $f_{\rm c}$ and varies with the cluster mass. Recent studies favour the interpretation that the ``missing'' baryons are 
located in the cluster outskirts so that the baryon fraction
reaches the cosmic value near the virial radius 
\cite{2010arXiv1007.1980R,2013arXiv1301.0624E}.
This behaviour is qualitatively reproduced only by the numerical simulations
that include either strong stellar feedback or active galactic nuclei (see Fig. 2 in \cite{2010MNRAS.405.2161D}). In these 
simulations, the dark-matter concentration (and therefore the dark-matter mass) of the clusters coincides with
what is measured in the corresponding N-body simulations 
to very good accuracy (see Fig. 8 in \cite{2010MNRAS.405.2161D}). 

Let us define a cluster as the spherical region centered on a density peak and
enclosing the overdensity $\Delta=500$ times the critical density of the 
universe.
Each cluster will be characterized by a well defined total mass $M_{\rm tot}$ 
(such that 
$M_{\rm tot}=M_{\rm b}+\tilde{M}_{\rm dm}$, where $\tilde{M}_{\rm dm}$ is the dark 
matter component of a cluster) and a baryon fraction 
$f_{\rm b}\equiv M_{\rm b}/M_{\rm tot}$. The same definition can be applied to identify haloes in a collisionless 
simulation. Note, however, that
particles in N-body simulations are assigned masses proportional to the total 
matter content of the Universe, 
i.e. $m_{\rm p}\propto \Omega_{\rm m}=\Omega_{\rm b}+\Omega_{\rm dm}$ (in terms of the baryonic and dark-matter components). 
Thus, by construction, the mass of simulated dark-matter haloes 
includes a baryonic fraction which coincides with $f_{\rm c}$. 
The total mass thus reads,
$M_{\rm nb}=M_{\rm dm}/(1-f_{\rm c})$, where $M_{\rm dm}$ is the 
corresponding dark matter mass. As discussed above, hydro simulations with strong feedback are the only ones which are able to reproduce the observed 
baryon distribution. Based on the fact that in these simulations the dark-matter mass of the clusters is hardly modified with respect to N-body simulations, in what follows we assume that 
$M_{\rm dm} =\tilde{M}_{\rm dm}$. The total mass of a cluster can be then linked with the total mass of the same halo in a N-body simulation $M_{\rm nb}$ via
\be\label{m}
M_{\rm tot}=\lp \frac{1-f_{\rm c}}{1-f_{\rm b}}\rp M_{\rm nb}.
\ee
Our assumption that $M_{\rm dm} =\tilde{M}_{\rm dm}$ should be seen as a working hypothesis supported by the match between
hydro simulations and observations. If future numerical simulations will evidence the need
of a more complicated treatment,
our analysis can be, of course, generalized by introducing a probability
density function $P(\tilde{M}_{\rm dm}|M_{\rm dm})$ at the expenses of 
introducing a number of extra parameters

\section{The observed mass function}

\begin{figure}[tbp]
\centering
\includegraphics[width=12cm]{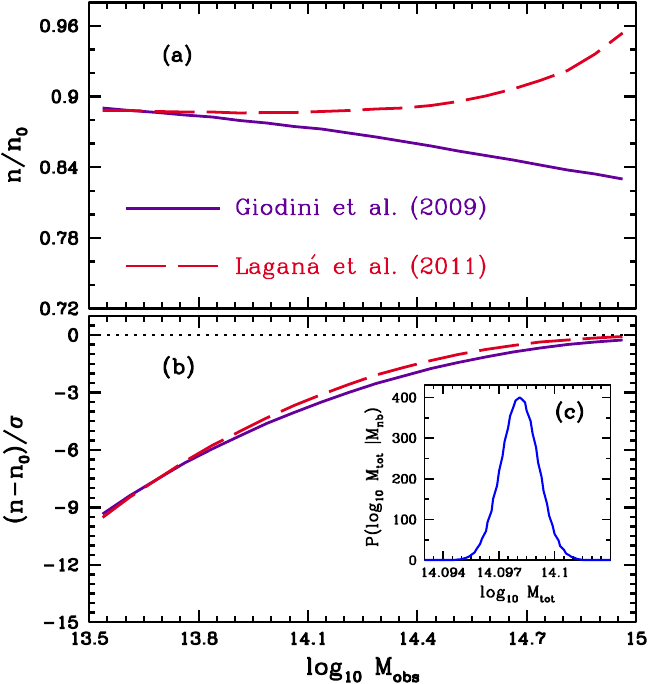}
\caption{(a) Ratio between the cluster mass function $n$ and
a model assuming that the baryonic mass fraction is always equal to the
cosmic value (i.e. $M_{\rm tot}=M_{\rm nb}$), $n_0$.
Line styles correspond to the observed baryon fraction reported by 
different authors. (b) Signal-to-noise ratio for the difference $n-n_0$ as a function
of the observed cluster mass in 20 equispaced log-bins. A survey volume 
of $0.8\, (h^{-1} {\rm Gpc})^{3}$ centered at $z=0.1$ is assumed. (c) The conditional probability density $\mP(\log_{10} M_{\rm tot}|M_{\rm nb})$ at a mass scale $M_{\rm nb}=1.3\times 10^{14}h^{-1}\,M_{\odot}$ as given in Eq.(\ref{prob2})} .
\label{massfun}
\end{figure}

Consider now a population of galaxy clusters and a set of haloes from a N-body 
simulation
both at the same redshift. Our goal is to write the mass function of the clusters in terms of that of the haloes (at a fixed $\Delta=500$). We need to keep into account that the baryon fraction of the real clusters is a stochastic 
quantity that varies from object to object. 
Based on observational studies, we assume that the conditional probability density
$P_{\rm b}(f_{\rm b}|M_{\rm tot})$, is well approximated by a Gaussian distribution with mean
\be\label{fa}
\langle f_{\rm b}|M_{\rm tot}\rangle =Af_{c} \lp \frac{M_{\rm tot}}{2\times 10^{14}\,h^{-1}M_{\odot}}\rp^{B},
\ee
and scatter $\sigma_{\rm b}$. 
We adopt the results by \cite{giodini},
$A=0.74\pm 0.01$, $B=0.09\pm 0.03$ and $\sigma_{\rm b} \simeq 2\times 10^{-3}$, as
fiducial values but it is worth stressing that other studies favour slightly different 
values for these parameters.
%
We want to write the conditional probability density $\mP(M_{\rm tot}|M_{\rm nb})$ which gives the 
distribution of the total cluster mass for a given $M_{\rm nb}$. 
This quantity satisfies the integral equation
\be\label{ie}
\mP(M_{\rm tot}|M_{\rm nb})=\frac{\rP(M_{\rm nb}|M_{\rm tot})}{n_{\rm h}(M_{\rm nb})}\int_{0}^{\infty}\!\!n_{\rm h}(x)\mP(M_{\rm tot}|x)\,\dd x,
\ee
where $\rP(M_{\rm nb}|M_{\rm tot})=(1-f_{\rm c})P_{b}(f_{\rm b}=f_{\star}|M_{\rm tot})/M_{\rm tot}$ with 
$f_{\star}\equiv 1-(1-f_{\rm c})M_{\rm nb}/M_{\rm tot}$. Since $\sigma_{\rm b}\ll \langle f_{\rm b}|M_{\rm tot}\rangle$, then $\mP(M_{\rm tot}|M_{\rm nb})$ 
is a narrow function centered at $\langle M_{\rm tot}|M_{\rm nb}\rangle \approx M_{\rm nb}(1-f_{\rm c})/(1-\langle f_{\rm b}|M_{\rm tot}\rangle)$, so that we can finally write
\be\label{prob2}
\mP(M_{\rm tot}|M_{\rm nb})= \frac{1-f_{\rm c}}{M_{\rm tot}}P_{b}(f_{\star}|M_{\rm tot})\frac{n_{\rm h}(\widetilde{M}_{\rm nb})}{n_{\rm h}(M_{\rm nb})}\mJ(M_{\rm tot}),
\ee
where $\widetilde{M}_{\rm nb}$ is the solution to the equation $M_{\rm tot}=\langle M_{\rm tot}|\widetilde{M}_{\rm nb}\rangle$ 
and \be 
\mJ(M_{\rm tot})=\left|\frac{\dd \langle M_{\rm tot}|M_{\rm nb}\rangle}{\dd M_{\rm nb}}\right|^{-1}_{M_{\rm nb}=\widetilde{M}_{\rm nb}}.
\ee 
The conditional probability density in Eq.(\ref{prob2}) represents the necessary tool to associate the 
masses of dark matter haloes defined in N-body simulations with the total masses of galaxy clusters in 
the presence of a mass dependent fraction of baryons. In general, 
$\mP(M_{\rm tot}|M_{\rm nb})$ is very well approximated by a log-normal distribution  with a mass-dependent 
log-scatter $\sigma_{\ln M_{\rm tot}|M_{\rm nb}}\approx \ln \left[1+\sigma_{\rm b}/(1-\langle f_{\rm b}|M_{\rm tot}\rangle)\right]$ 
(see panel (c) of Fig. ~\ref{massfun} for an example).

Cluster masses are not observed directly and need to be inferred from
observational proxies (e.g. optical richness, X-ray temperature or flux, 
Sunyaev-Zel'dovich signal, lensing shear). 
We assume that the observed mass, $M_{\rm obs}$, is an unbiased estimate of the 
total mass, with a log-normal probability density function of the residuals $P(M_{\rm obs}|M_{\rm tot})$
\cite{lima,cunha}, 
for which we use a constant log-scatter $\sigma_{\ln M_{\rm obs}|M_{\rm tot}}=0.1$. 
In general the scaling relation $P(M_{\rm obs}|M_{\rm tot})$ is obtained observationally, 
and for this reason there is no need to explicitly emphasize its dependence on
the baryonic fraction (e.g, the X-ray luminosity depends on the total amount of gas) in our formalism. 
The mass function of galaxy clusters in terms of their observed mass is then
\be\label{nX}
n(M_{\rm obs},z)=\int_{0}^{\infty}n_{\rm h}(M_{\rm nb},z)\,P(M_{\rm obs}|M_{\rm nb})\, \dd M_{\rm nb},
\ee
with 
\be
P(M_{\rm obs}|M_{\rm nb})=\int_{0}^{\infty} P(M_{\rm obs}|M_{\rm tot})\mP(M_{\rm tot}|M_{\rm nb})\,\dd M_{\rm tot}.
\label{probabilities}\ee 
Note that the conditional probability $P(M_{\rm obs}|M_{\rm nb})$ can be written as a log-normal distribution with scatter $[\sigma^{2}_{\ln M_{\rm obs}|M_{\rm tot}}+\sigma^{2}_{\ln M_{\rm tot}|M_{\rm nb}}]^{1/2}$. The main effect of the variable baryon fraction is therefore to introduce a systematic mass-dependent offset between $M_{\rm obs}$ and $M_{\rm nb}$ while the correction to the intrinsic scatter plays a sub-dominant role, at least for mass proxies with broad distributions ($\sigma_{\ln M_{\rm tot}|M_{\rm nb}}\ll \sigma_{\ln M_{\rm obs}|M_{\rm tot}}$).

In panel (a) of Fig.~\ref{massfun} we compare the cluster mass function, 
$n(M_{\rm obs})$, with its counterpart obtained 
by neglecting the effects of the
varying baryon fraction (i.e. assuming that $M_{\rm tot}=M_{\rm nb}$) 
that we dub $n_0(M_{\rm obs})$ - note that $n_0$ differs from $n_{\rm h}$ due
to the mass-measurement errors. We use the observational scaling relations $P_{\rm b}(f_{\rm b}|M_{\rm tot})$ by \cite{giodini} and \cite{lagana}. The latter corresponds to Eq.~(\ref{fa}) with $A\simeq 0.8$ and $B=0.136\pm 0.028$.
The main effect is a reduction of the 
cluster counts by $\sim 5-15\%$, depending on the cluster mass and
the details of $P_{\rm b}(f_{\rm b}|M_{\rm tot})$.

Will the discrepancy between the actual cluster mass function, $n(M_{\rm obs})$, 
and the predictions of N-body simulations (convolved with the
mass-measurement error), $n_0(M_{\rm obs})$, be 
noticeable with future observational campaigns?
As a prototype for the forthcoming cluster surveys, 
we consider a catalog spanning
a comoving volume of $0.8\, (h^{-1}{\rm Gpc})^{3}$ 
(corresponding to a survey covering the full sky down to $z<0.2$ or half of the
sky to $z<0.25$)
and containing
$\sim 2.79\times 10^{4}$ entries in the mass range 
$10^{13.5}<M_{\rm obs}/( h^{-1} M_{\odot})<10^{15}$ (for our fiducial model)
with mean redshift $z=0.1$. We assume full completeness in the mass range we are probing and
compute the cluster mass function in $20$ 
mass bins of width $\Delta_{\log_{10} M_{\rm obs}}=0.075$.
Assuming Poisson errorbars of size $\sigma$, in panel (b) of Fig.~\ref{massfun} we show the signal-to-noise ratio
for the difference $n(M_{\rm obs})-n_0(M_{\rm obs})$ as a function of the cluster
mass. 
Highly statistically significant deviations are detectable for $M_{\rm obs}<
10^{14.5} h^{-1}M_\odot$.

\begin{table}[t]
\centering
\scalebox{0.78}{%
\begin{tabular}{|c|c|c|c|c|} \hline
Case & Model & Notes & Data & Description \\
\hline
I&$n_0$ & & $n_{\rm obs}$& The halo mass function from N-body simulations is used to fit the observed mass function\\
II&$n_0$&$\uparrow$ cov& $n_{\rm obs}$& 
As in I but after artificially
inflating the covariance matrix of the $n_{\rm h}$ parameters \\
III & $n$ & &  $n_{\rm obs}$ &
Accounting for a varying baryon fraction
and simultaneously fitting $P(f_{\rm b}|M_{\rm tot})$\\
IV & $n$ & & $n_{\rm obs}, f_{\rm b}$ & As in III combining the mass function data with 30 measurements of $f_{\rm b}$\\
V &  $n$ & &$n_{\rm obs}$ & As in III but assuming perfect a priori knowledge of $P(f_{\rm b}|M_{\rm tot})$\\
\hline
I$_{\rm G}$ & $n_0$ & &  $n_{\rm obs}$& As in I marginalizing
over the mass-measurement error with a Gaussian prior (see text)\\
I$_{\rm F}$ & $n_0$ & &  $n_{\rm obs}$& As in I marginalizing
over the mass-measurement error with a broad flat prior (see text)\\
IV$_{\rm G}$ & $n$ & & $n_{\rm obs}, f_{\rm b}$ & As in IV 
marginalizing over the mass-measurement error with a Gaussian prior (see text)\\
IV$_{\rm F}$ &  $n$ & & $n_{\rm obs}, f_{\rm b}$ & As in IV 
marginalizing over the mass-measurement error with a broad flat prior (see text)\\
\hline
\end{tabular}
}
\caption{\label{tablemodels} Summary of the Bayesian-inference cases discussed in this paper. Note that $n_0$ is obtained assuming that all clusters have
$M_{\rm tot}\equiv M_{\rm nb}$ (i.e. $f_{\rm b}=f_{\rm c}$) in Eq.~(\ref{probabilities}) while $n$ takes into account that the baryonic mass fraction varies from object to object.}
\end{table}

\section{Cosmological parameters}\label{sec:theory}

\begin{figure}[t]
 \includegraphics[width=15.5cm]{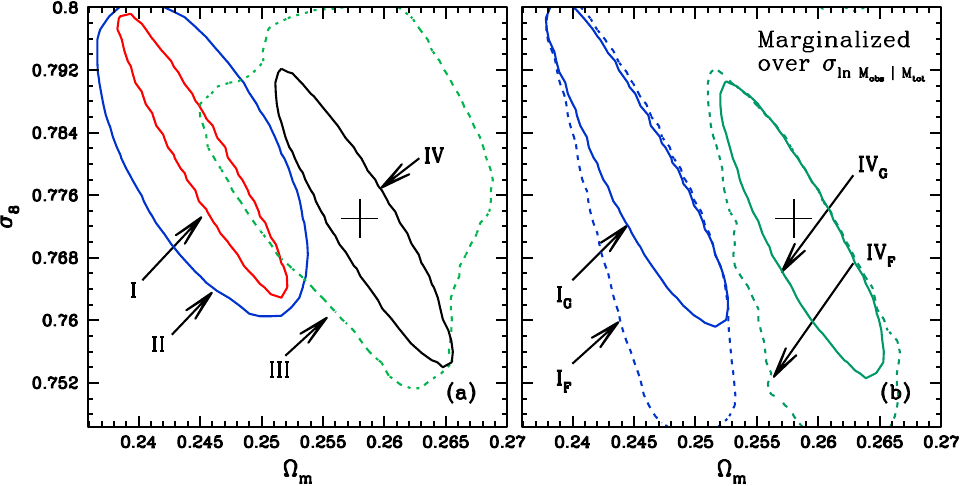}
\caption{(a) The $95.4\%$ credibility region in the plane
$\{\Omega_{\rm m}, \sigma_{8}\}$ for the cases discussed in Table \ref{tablemodels}. 
A cross indicates the input values for the cosmological parameters. 
(b) As in panel (a) but treating the measurement error 
$\sigma_{\ln M_{\rm obs}|M_{\rm tot}}$ as a nuisance parameter 
(contrary to panel (a) where the log-scatter is assumed to be known exactly).}
\label{post}
\end{figure}

We want to quantify the bias and the uncertainty in the measurement of the 
cosmological parameters $\Omega_{\rm m}$ and $\sigma_{8}$ obtained 
by fitting the observed cluster mass function, $n_{\rm obs}$, with different models.
In order to do this, we use the mock cluster catalog
described in the previous section 
and we sample the posterior distribution of the model parameters
with a Markov Chain Monte Carlo algorithm. 
We use different models and combinations of data which are briefly summarized 
in Table \ref{tablemodels} and extensively described below.
The (marginalized) posterior mean and rms values 
for $\Omega_{\rm m}$ and $\sigma_{8}$ are given in  Table \ref{tableres}
together with the corresponding ``figure of merit'' 
(FoM, defined as the inverse of the area of the joint $68.3\%$ credibility 
region in the $\{\Omega_{\rm m},\sigma_{8}\}$ plane).
The joint $95.4\%$ credibility regions are instead
shown in panel (a) of Fig.~\ref{post}. 

First, we consider the mass function extracted from
N-body simulations with no corrections for a varying $f_{\rm b}$ (case I).
Specifically, we use the function $n_0(M_{\rm obs})$ to fit the observed mass
distribution.
The resulting estimate for $\Omega_{\rm m}$ 
is significantly biased low. 
To first order, this is because
the normalization of the function $n_0$ scales proportionally to $\Omega_{\rm m}$
and, as shown in panel (a) of Fig.~\ref{massfun}, the effect of the varying
baryon fraction is to reduce the overall normalization of the observed
cluster counts.
On the other hand, $\sigma_8$ is slightly biased high (for our fiducial model)
as expected from the location of the exponential cutoff in the
mass function. However, the bias is not very significant given the
corresponding statistical uncertainty.
The size and sign of the bias 
on $\sigma_8$ markedly depends
on the values adopted in Eq.~(\ref{fa}) while this is not true
for $\Omega_{\rm m}$, which is always biased low for $A<1$.

To gain freedom in the shape of the theoretical mass function
and reduce systematic effects when fitting current data,
it is common practice to let the parameters that define $n_{\rm h}$ 
vary within some predefined range \cite{mantzI}.
We have investigated what happens applying this technique to our mock sample.
In this case we have allowed the parameters of the mass function to vary
within a four-dimensional Gaussian prior. 
We built the covariance matrix of the prior by
multiplying the original covariance matrix of the parameters  
(kindly made available by Jeremy Tinker) with a positive constant so that
the halo mass function at $M=10^{15}\, h^{-1} M_\odot$ is $\sim 10\%$ uncertain (case II).
With respect to case I, this method does not improve the bias of $\Omega_{\rm m}$ and $\sigma_{8}$
while the corresponding FoM decreases by a factor of $\sim 3.6$.
\begin{table}[t]
\center
\begin{tabular}{|c|c|c|c|} \hline
Case & $\Omega_{\rm m}$  & $\sigma_{8}$& FoM  \\
\hline
Fiducial & $0.258$ &  0.773 & \\
I &  $0.245 \pm 0.002$ & $0.780 \pm 0.007$& $2.4\times 10^{4}$  \\
II &  $0.244\pm 0.003$ & $0.780\pm 0.007$&  $6.7\times 10^{3}$ \\
III &  $0.259\pm 0.005$ &  $0.782\pm 0.012$& $2.8\times 10^{3}$ \\
IV & $0.257\pm 0.003$  & $0.773\pm 0.007$& $2.1\times 10^{4}$ \\
V &  $0.258\pm 0.003$ &  $0.772\pm0.007$& $2.5\times 10^{4}$ \\
\hline 
\end{tabular}\caption{\label{tableres}
Mean and rms value of the marginalized posterior distribution for 
$\Omega_{\rm m}$ and $\sigma_8$. The third column shows the figure-of-merit, defined as the 
inverse of the area of the joint $68.3\%$ credibility region.
}
\label{table}
\end{table}

In order to eliminate the bias, we release the assumption that
$M_{\rm tot}=M_{\rm nb}$ in Eq.~(\ref{probabilities}) and
replace $n_{0}$ with $n$ to fit $n_{\rm obs}$ (case III). This way, we simultaneously constrain 
the cosmological parameters and the scaling relation $P(f_{\rm b}|M_{\rm tot})$.
This procedure is analogous to the ``self calibration'' method proposed
to extract cosmological information from cluster surveys 
\cite{hu}. We have verified that this approach provides
unbiased estimates of $\Omega_{\rm m}$ and $\sigma_{8}$. However, 
the statistical uncertainty on the values of the cosmological parameters
constrained by the cluster counts
are significantly larger with respect to case I and II.

The situation markedly improves 
by considering additional information on $f_{\rm b}$
extracted from multi-wavelength studies of a small subset of
galaxy clusters. 
To show this, we randomly select $30$ clusters 
out of the full sample 
and imagine that the baryonic fraction of their mass content
has been measured with $10\%$ precision in an unbiased way (case IV).
We then build new Markov chains assuming that the 
information from the measurement of the baryon fraction is independent
of the cluster counts.
The resulting
estimates for $\Omega_{\rm m}$ and $\sigma_{8}$ 
are unbiased and errorbars small as expected
from experiments for ``precision cosmology''. In this case, 
the parameters of the scaling relation $P(f_{\rm b}|M_{\rm tot})$ are also recovered to good accuracy, namely (to $68.3\%$ credibility), $A=0.74\pm 0.01$, $B=0.09\pm 0.01$ and $\sigma_{\rm b}=4.5^{+3.8}_{-3.1}\times 10^{-3}$.

Finally, we consider the ideal case in which the scaling relation
$P(f_{\rm b}|M_{\rm tot})$ is perfectly known from independent data (case V,
not shown in Fig.~\ref{post}).
This allows us to conclude that
case IV gives cosmological constraints which are nearly optimal.

To simplify the discussion, so far, we have assumed that the scatter of
the observational mass estimates for the galaxy clusters, 
$\sigma_{\ln M_{\rm obs}|M_{\rm nb}}$,
is perfectly known. This, however, does not accurately reflect reality where
one has only limited knowledge on the actual size of the measurement error
which should then be treated as a nuisance parameter in the model-fitting 
procedure \cite{cunha}.
This will generally broaden the credibility region of the cosmological 
parameters and, in principle, might reduce the statistical significance of
the biased estimates obtained with model $n_0$. 
In order to evaluate the impact of this subtlety on our results, 
we have repeated the measurement of the cosmological parameters
(for case I and IV)
after marginalizing over $\sigma_{\ln M_{\rm obs}|M_{\rm nb}}$.
We have considered two cases. First, as a realistic option, we have adopted 
a Gaussian prior on $\sigma_{\ln M_{\rm obs}|M_{\rm nb}}$ 
with mean 0.1 and rms error 0.02 (cases I$_{\rm G}$ and IV$_{\rm G}$) -
note that a $20\%$ standard error of the standard 
deviation corresponds to a sample of 14 objects.
As a (very) pessimistic option, instead, we have considered
a flat prior between $0<\sigma_{\ln M_{\rm obs}|M_{\rm nb}}<1$ (cases I$_{\rm F}$ and
IV$_{\rm F}$). 
The corresponding joint posterior distributions for 
$\Omega_{\rm m}$ and $\sigma_8$
are shown in panel (b) of Fig.~\ref{post}. In all cases, the bias in the
estimates based on the model $n_0$ is still evident.

\section{Discussion and conclusions}\label{sec:results}

The fractional baryon content of galaxy clusters within an overdensity of $\Delta=500$ is observed to be a random variable which, on average, decreases with the cluster mass \cite{giodini,lagana,lin,sun}. 
Therefore the cluster mass function must differ from the predictions
of N-body simulations where the baryon fraction is implicitly held constant
to the cosmic value.

The forthcoming generation of cluster surveys will provide number counts 
with an accuracy ranging between $1$ and $10\%$ depending on the cluster mass.
Such an error is substantally smaller than the difference between models of the mass function
with and without baryonic physics.
Considering a prototypic catalog of galaxy clusters 
 containing $2.79\times 10^4$ entries,
we have shown that constraints on the cosmological parameters $\Omega_{\rm m}$
and $\sigma_{8}$ derived from the cluster mass function 
would be severely biased if this signal is modelled 
with fitting formulae based on N-body simulations of collisionless matter. 
In particular, $\Omega_{\rm m}$ would be always biased low while
the bias on $\sigma_8$ depends on the details of the scaling relations
between $f_{\rm b}$ and the cluster mass. 

The widespread technique of artificially inflating the covariance matrix of the parameters that describe the halo mass function to gain freedom and minimize systematic effect will not be of much help.
Our study shows that it would enlarge the statistical uncertainties on the cosmological parameters
without eliminating (or reducing) the bias.

In order to obtain accurate estimates for the cosmological parameters, 
complementary information on the fraction of baryons as a function of the total mass is 
required. The optimal method to eliminate this systematic effect requires
two ingredients: I) an accurate model for the conditional probability density
of finding a particular value for
$f_{\rm b}$ given the cluster total mass, $P(f_{\rm b}|M_{\rm tot})$; 
II) A small, random subsample of clusters with follow-up data for which 
simultaneous measurements of $f_{\rm b}$ and $M_{\rm tot}$ can be made.
We have shown that, if the scatter around the mean $f_{\rm b}-M_{\rm tot}$
relation, $\sigma_{\rm b}$, is independent of mass, nearly 30 objects would be
enough for precision cosmology.
The required size of the subsample should grow bigger if 
$\sigma_{\rm b}$ has a strong mass dependence.

It is interesting to compare the precision we can achieve with the different methods in our fiducial case (see Table \ref{tableres}).
Using the N-body mass function (case I) returns estimates for and $\sigma_8$ and $\Omega_{\rm m}$ with statistical errors of 1\% and 2\%, respectively (but with a systematic shift which is approximately $1$ and $6.5$ times larger).
On the other hand, accounting for the variable baryon fraction (case III) eliminates the biases but
nearly doubles the rms of the marginal probabilities due to the inclusion of three additional parameters.
However, combining this method with sufficient follow-up information (case IV), we can optimally recover the same uncertainties as in case I.

Let us now critically discuss the formalism we have outlined in this Paper.
By performing an object by object comparison between collisionless and hydrodynamic simulations with the 
same initial conditions,
recent studies have shown that the dark-matter mass of a cluster (within an overdensity of $\Delta=500$) 
does not change when baryonic physics is included, provided a strong form of feedback
is considered. Our analysis is based on this result. This is a conservative assumption which, if violated, would introduce additional deviations in the cluster mass function and thus make estimates of the cosmological parameters based on the standard approach even more biased (since it is very unlikely that this effect would exactly cancel the mass dependence of $f_{\rm b}$). However, in Section 2, we have indicated how our calculations could be easily generalized to that case. Independent from the detailed origin of the mass discrepancy,
the take-home message of this Paper is that baryonic physics will have a sizable impact on the measurement
 of cosmological parameters from cluster number counts.

To facilitate understanding, our simple analysis considers a complete sample in a narrow redshift bin 
and only two cosmological parameters but our results are of general value. 
It is straightforward to generalize them including more parameters and 
accounting for possible evolutionary
effects in the baryon fraction along the past light cone 
and for the radial selection function of a realistic survey.
This, however, is beyond the scope of this Paper.

We conclude that considering
the effect of baryons on the cluster mass function 
is central to extract unbiased estimates of the cosmological parameters 
from forthcoming large-volume surveys such as DES \cite{abbott},
\emph{eROSITA} \cite{predehl, pillepich}, ASKAP-EMU \cite{norris},
CCAT \cite{radford}, LSST \cite{abell} and Euclid \cite{laureijs}.

\acknowledgments
We acknowledge support through the SFB-Transregio 33 ``The Dark Universe'' by the 
Deutsche Forschungsgemeinschaft (DFG).

\bibliographystyle{JHEP}	
\bibliography{refs}		
\end{document}